\documentclass[ 
 amsmath,amssymb,
 aps,
prb,twocolumn,
dvipdfmx
]{revtex4-1}

\usepackage{graphicx}
\usepackage{dcolumn}
\usepackage{bm,bbm}

\usepackage{caption}
\usepackage{amsmath}
 \usepackage{amssymb}
 \usepackage{amsfonts}
\usepackage{relsize}
\usepackage{braket}
\usepackage{ulem}
\usepackage{here}
\usepackage{color}
\usepackage{comment}

\newcommand{\bq}{ \bm{q} }
\newcommand{\QQ}{ \bm{Q} }

\usepackage{hyperref}
\usepackage{url}
\allowdisplaybreaks[4]

\begin{document}


\title{Electronic origin of ferroic quadrupole moment under antiferroic quadrupole orders and finite magnetic moment in $J_{\rm eff}=3/2$ systems 
}

\author{Haruhiro Kubo}
\author{Takayuki Ishitobi}

\author{Kazumasa Hattori}%
\affiliation{%
 Department of Physics, Tokyo Metropolitan University,\\ 1-1, Minami-osawa, Hachioji, Tokyo 192-0397, Japan
}%

\date{\today}

\begin{abstract}
We study the electronic origin of parasitic ferroic 
quadrupole moments in antiferroic quadrupole orders  
by extending a model studied in G. Chen et al., 
Phys. Rev. B {\bf 82}, 174440 (2010) 
with the effective angular momentum 
$J_{\rm eff}=3/2$ quartet ground states. Taking into account the 
first crystalline-electric-field (CEF) excited doublet, 
cubic anisotropy in the 
quadrupole moments emerges, which leads to the 
induced ferroic quadrupole moments in the 
antiferro quadrupolar phases. The hybridization 
with the CEF excited  quartet states also causes  finite magnetic 
moments compatible to the observed size of the 
effective moment in typical $J_{\rm eff}=3/2$ systems, as opposed to the naive 
expectation of vanishing moments in the 
$J_{\rm eff}=3/2$ systems. These results suggest 
the importance of the corrections arising from the high-energy CEF excited states in the $J_{\rm eff}=3/2$ systems.

\end{abstract}

\pacs{Valid PACS appear here}
\maketitle


\section{Introduction}
Strongly correlated electrons systems with orbital degrees of freedom have attracted great attention in recent years \cite{Tokura2000}. The orbital degrees of freedom possess potential functions alternative to the modern magnetic-based devices\cite{Fiebig2016-or}. Combining the orbital and conventional spin degrees of freedom inevitably leads to the notion of multipole moments\cite{Kuramoto2009-ll}. Such multipole moments play important role in correlated electron systems with strong spin-orbit couplings.  They can exhibit various fascinating phenomena such as topological spin-orbital Mott insulators\cite{Pesin2010-py}, spin-orbital liquid states\cite{Jackeli2009,nasu2014vaporization,Corboz2012-ei,Kitagawa2018-yj}, superconductivity mediated by multipolar spin-orbital fluctuations\cite{Nomoto2016-nd,Hattori2017-rg}, and in doped Mott insulators\cite{Watanabe2013-ap}.

For activating multipole physics, 
one needs high symmetry and strong 
spin-orbit couplings. Possible candidates 
are Mott insulators with $d^1$ 
configuration surrounded by a regular 
oxygen octahedron. Although materials 
with $3d$ electrons often exhibit 
the Jahn-Teller distortion as 
lowering temperature, some 5$d$ 
electron systems keep their cubic 
symmetry even at low temperatures. 
As a such $d$ electron system, 
double-perovskite compounds 
$A_2B$$Tr$O$_6$ ($A$=Ba, Sr,Ca; $B$= 
Mg, Ca, Sr, Ba, Zn, Cd, $Tr=$Re, Os, Mo)\cite{Vasala2015-ab,
Longo1961-fm,Wiebe2003-nd, 
doi:10.1021/ic50002a010,Chamberland1979-yu,
Abakumov1997-qi,Wiebe2002-tn,Wiebe2003-nd,
Bramnik2003-qg,Yamamura2006-sq,
Marjerrison2016-ww,Hirai2019-en} 
and Ta chlorides $A_2$TaCl$_6$ ($A=$K, Rb, 
Cs)\cite{Ishikawa2019-xg} have been 
recently studied 
intensively. 
The oxygen 
octahedron crystalline-electric-field (CEF) lifts 
the ten-fold degeneracy of the $d$ electrons to the high-energy 
$e_g$ and the low-energy $t_{2g}$ states. 
The latter is split further by the 
spin-orbit coupling and form 
so-called effective angular 
momentum $J_{\rm eff}=3/2$ ground 
state and $J_{\rm eff}=1/2$ first 
excited state. In these compounds, 
the spin-orbit coupling $\lambda$ is of 
the order of $\sim$0.2--0.4 eV, while 
the CEF gap between 
the ground $t_{2g}$ and the excited 
$e_g$ states $D$ is $D\sim$3--5 eV \cite{Paramekanti2018-pb,Oh2018-sn,Ishikawa2019-xg}.

What is unique to these systems is that the ground state with $J_{\rm eff}=3/2$ has no magnetic dipole moment as a result of exact cancellation between the spin and orbital angular momenta. This leads to various possibilities of multipole orders at low temperatures. For example, Re based double-perovskites exhibit ferro and antiferro quadrupole orders in addition to dipole-octupole magnetic orders. In Ba$_2$$Tr$OsO$_6$ ($Tr=$Zn, Mg, and Ca) with $d^2$ configuration, octupole orders have been suggested recently.\cite{Paramekanti2020-ae,Maharaj2020-pk} 

The vanishing magnetic dipole moment is 
also reflected in the small moment typically 
$\sim 0.3$--$0.8\mu_{\rm B}$ in 
their high-temperature magnetic 
susceptibility, where $\mu_{\rm B}$ is 
the Bohr magneton. This small but 
finite value has been interpreted as a 
partial cancellation of the spin and 
orbital angular momenta owing to 
the hybridization between the $d$ and 
the oxygen $p$ orbitals.\cite{Ahn2017-hi,Hirai2019-en,Ishikawa2019-xg}

In the pioneering work by Chen et al.,\cite{Chen2010} a model containing the ground-state CEF quartet originating from the $t_{2g}$ orbital was constructed in the limit of the strong spin-orbit coupling $\lambda\to \infty$. The quartet can be described by the effective angular momentum $J_{\rm eff}=3/2$ and they predicted several interesting multipolar ordered states in terms of the multipole moments of the $J_{\rm eff}=3/2$ multiplet: an antiferroic quadrupole order with the $e_g$ irreducible representation and ferro- and antiferro-magnetic 
dipole-octupole orders. These results are indeed supported by the experimental observation of the phase transitions e.g., in Ba$_2$MgReO$_6$\cite{Hirai2019-en} and Ba$_2$NaOsO$_6$.\cite{Lu2017}  

Recently, Hirai et al., reported that ferroic quadrupole moments $\sim O_{20}=3z^2-r^2$ emerge under the antiferroic order of the type $O_{22}=x^2-y^2$ below $T_q=33$ K by their x-ray experiments\cite{Hirai2020}. The presence of the ferroic moment can be understood by a simple symmetry argument; the free energy for the $e_g$ orbital moments contains a cubic coupling $\sim O_{22}^2O_{20}$. This induces ferroic quadrupole moments proportional to the square of the antiferroic ones. They discuss this can be due to the Jahn-Teller effects and the lattice anharmonicity. The former has been recently analyzed and successfully reproduced the emergence of the ferroic quadrupole moments\cite{Iwahara2022-mt}.

In this paper, we point out that 
$d$ electron CEF excited states  
can influence the CEF ground quartet 
with $J_{\rm eff}=3/2$, 
which has no magnetic dipole moment 
and no cubic coupling 
$\sim O_{22}^2O_{20}$ without couplings to other degrees of freedom.  
In the $J_{\rm eff}=3/2$ multiplets, 
the local cubic anisotropy which arises from the CEF potential 
vanishes. The anisotropy can emerge 
when the CEF excited states are taken into account. Such anisotropy 
due to the excited spin-singlet state ($\Gamma_1$ in the cubic 
symmetry) are indeed discussed in the $\Gamma_3$ non-Kramers 
doublet ground state in Pr-based compounds\cite{Hattori2014,Hattori2016,
Ishitobi2021,Tsunetsugu2021,Hattori2022-ss} and the analysis 
there is also applicable to the case of the $d^1$ systems. 
This is because the $J_{\rm eff}=3/2$ states is classified as 
$\Gamma_8$ irreducible representation (irrep) in the cubic symmetry and can be regarded as 
a product of the spin-1/2 and the orbital $e_g(\Gamma_3)$. 
Indeed, the CEF first-excited state is the $\Gamma_7$ state 
which is the spin-1/2 and the orbital-singlet state. Thus, 
apart from the spin degrees of freedom, the orbital sector 
is identical in the two systems. For the small but finite magnetic dipole moment, it will be shown that the corrections of order $\lambda/D\sim 0.1$ leads to non-negligible contribution to the magnetic moment in realistic systems in this paper.

This paper is organized as follows. In Sec.~\ref{sec:model}, we review the model proposed in Ref.~\onlinecite{Chen2010} and explicitly introduce the matrix form of quadrupole operators including the excited states. The exchange Hamiltonian is rewritten in terms of these quadrupole and spin-orbital operators to make this paper self-contained form. In Sec.~\ref{sec:results}, we show the results of the two-site mean-field approximation. We discuss the effects of the excited CEF state on the ferro components of the order parameter in the antiferroic quadrupole  ordered state and also on the phase-diagram. The temperature-magnetic field phase diagrams are also analyzed. In Sec.~\ref{sec:dis}, we discuss the results in this paper and related materials. We also discuss how the  finite magnetic moment emerges in the ground state $J_{\rm eff}=3/2$ state. Finally,  Sec.~\ref{sec:sum} summarizes this paper.


\section{Model}\label{sec:model}
We start by introducing the general model describing $d^1$ electron configuration for arbitrary spin-orbit coupling and CEF strength. This means that both $t_{2g}$ and $e_g$ degrees of freedom together with the spin $1/2$ ones are taken into account. We then derive an effective $t_{2g}$-dominant model with six states, ignoring the excited $e_g$ dominant states. The interaction between the effective $t_{2g}$ electrons are introduced as similarly to the study in Ref.~\onlinecite{Chen2010}.  We will not project them onto the effective total angular momentum $J_{\rm eff}=3/2$, but keep both $J_{\rm eff}=3/2$ and $1/2$ constructed by the $t_{2g}$ orbitals. 

\subsection{Local Hamiltonian}
The local part of the Hamiltonian is a conventional one with the CEF parameter $B_4$ and the spin-orbit coupling $\lambda>0$ as 
\begin{align}
	H_{\rm loc} &= H_{\rm CEF}+H_{\rm SO}, \label{eq:Hloc}\\
	H_{\rm CEF}&=\frac{B_4}{2}\sum_{i,\sigma,m,m'} \gamma_{mm'}d_{m\sigma}^\dag d_{m'\sigma},\\
	H_{\rm SO}&= \lambda \sum_{i,\sigma,m,m'}  ({\bm L})_{mm'}\cdot \bm{s}_{\sigma\sigma'}d_{m\sigma}^\dag d_{m'\sigma'},
\end{align}
Here, $d_{m\sigma}$ is the $d$ electron annihilation operator with the $z$ component of the orbital angular momentum $m=0,\pm 1, \pm 2$ and the spin $\sigma=\uparrow,\downarrow$. The CEF potential for the $d$ electrons are parameterized by $\gamma_{mm'}\equiv (\frac{15}{2}m^2 -\frac{35}{2}|m|+6)\delta_{m,m'}+5\delta_{|m-m'|,4}$, where $\delta_{m,m'}$ is the Kronecker delta.
$\bm{L}$ and $\bm s$ are the orbital angular momentum and the spin-1/2 matrices for the $d$ electrons, respectively.

The eigenstates of $H_{\rm loc}$ are split into $4+2+4$ and their eigenvalues are given as 
\begin{align}
	\epsilon_{1/2}&=-2 B_4+\lambda, \label{eq:e12}\\
	\epsilon_{3/2}&=\frac{1}{4} \left(2 B_4-\lambda -\sqrt{5} \sqrt{20 B_4^2+4 B_4 \lambda +5 \lambda ^2}\right),\\
	\epsilon_{3/2}'&=\frac{1}{4} \left(2 B_4-\lambda +\sqrt{5} \sqrt{20 B_4^2+4 B_4 \lambda +5 \lambda ^2} \right),
\end{align}
where the subscripts represent the effective angular momenta; $3/2$ ($1/2$) corresponds to $\Gamma_8$($\Gamma_7$) state in the cubic point group O$_h$. 

For small $\lambda/B_4$ with $B_4>0$, 
\begin{align}
	\epsilon_{3/2}&\simeq -2 B_4-\frac{\lambda }{2}-\frac{3 \lambda ^2}{10 B_4}+\cdots, \label{eq:e32a}\\
	\epsilon_{3/2}'&\simeq 3B_4+\frac{3 \lambda ^2}{10 B_4}+\cdots.
	\label{eq:e32b}
\end{align}
When the CEF potential is sufficiently large, $\epsilon_{3/2}'$ is much larger than the other two and the excitation gap $D$ is $D\simeq 5B_4\sim 3$--$5$ eV for the compounds mentioned in the Introduction.\cite{Paramekanti2018-pb,Oh2018-sn,Ishikawa2019-xg} Thus, the quartet with its eigenenergy $\epsilon_{3/2}$ corresponds to the ground state with $J_{\rm eff}=3/2$ and the first excited state is that with $\epsilon_{1/2}$. The former approximately consists of the $t_{2g}$ electrons, while the latter is purely $t_{2g}$ origin, see Appendix~\ref{app:wavefunc}. In the following, we will concentrate on these six states and ignore the higher energy states at $\epsilon_{3/2}'$. The matrices which will be shown in the following [Eqs.~(\ref{eq:matJz}),(\ref{eq:matu}), and (\ref{eq:matv})] are calculated within the full ten-dimensional Hilbert space with keeping $O(\varepsilon^2)$ terms for later purposes. The six states can be labeled by the diagonal elements of the total angular momentum $J_z$ for small $\varepsilon\equiv \lambda/B_4\sim 0.5$:

\begin{align}
	J_z &\simeq \frac{1}{3}\begin{bmatrix}
		\frac{5}{2} & 0            & -2\sqrt{2}+\alpha_J & 0 & 0 & 0\\
		           & -\frac{5}{2} & 0 & -2\sqrt{2}+\alpha_J & 0 & 0\\		
		           &             & \frac{1}{2}+\beta_J & 0 & 0 & 0\\		
		           &  & 0 & -\frac{1}{2}-\beta_J & 0 & 0\\		
		           &             &  &  & -\frac{3}{2} & 0\\		
		           &             &  &  &  & \frac{3}{2}\\				
	\end{bmatrix}. \label{eq:matJz}
\end{align}
Here the empty parts in the matrix have been omitted since $J_z=J_z^\dag$. 
The constants $\alpha_J$ and $\beta_J$ are given as 
$
	\alpha_J=\frac{3\sqrt{2}}{5}\varepsilon+O\left(\varepsilon^3\right),\quad 
	\beta_J=\frac{12}{5}\varepsilon-\frac{9}{25}\varepsilon^2+O\left(\varepsilon^3\right).
$
Note that the factor $1/3$ in Eq.~(\ref{eq:matJz}) and the basis of the matrix $J_z$ is the eigenstates for $H_{\rm loc}={\rm diag}(\epsilon_{1/2},\epsilon_{1/2},\epsilon_{3/2},\epsilon_{3/2},\epsilon_{3/2},\epsilon_{3/2})$. See the wavefunctions in Appendix~\ref{app:wavefunc}. From this expression, it is clear that the quartet can be described by the effective angular momentum $J_{\rm eff}=3/2$ with $J_{\rm eff}^z=3J_z$ when the excited states are ignored for $\lambda\to \infty$ with $\varepsilon=\lambda/B_4\to 0$. 

\begin{figure}[t!]
\begin{center}
\includegraphics[width=0.4\textwidth]{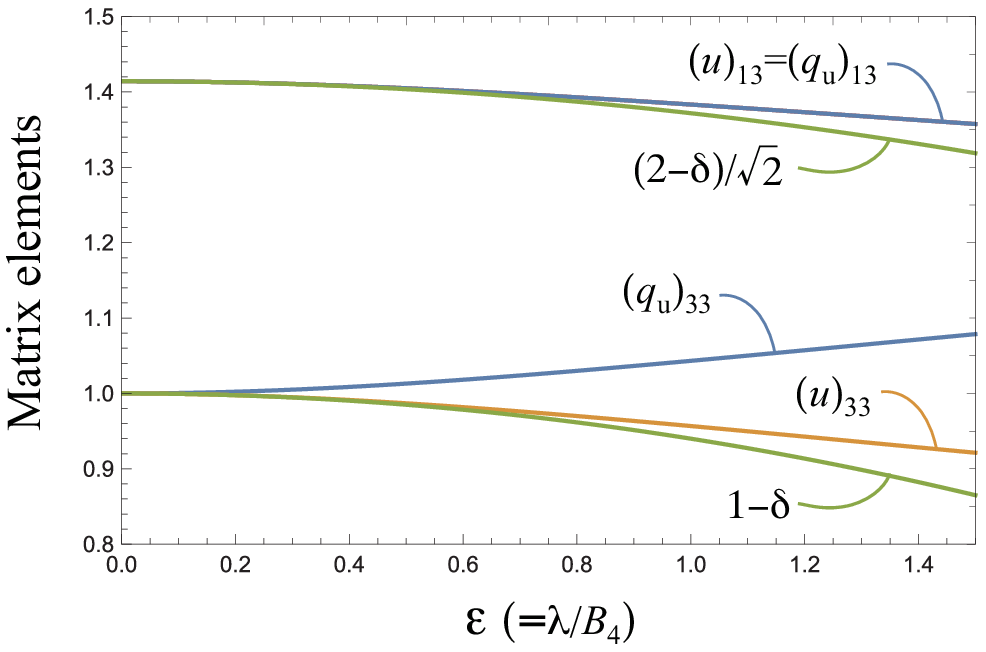}
\end{center}
\caption{The matrix elements of $u$ and $q_u$ as a function of $\varepsilon=\lambda/B_4$. 
For the offdiagonal elements, $u$ and $q_u$ are identical: $(u)_{13} = - (u)_{24}=(q_{u})_{13}=-(q_{u})_{24}$, while
for the diagonal ones, they are different for finite $\varepsilon$: $(u)_{33}=(u)_{44}=-(u)_{55}=-(u)_{66}$.
$(q_u)_{33}=(q_u)_{44}=-(q_u)_{55}=-(q_u)_{66}$. The approximated values for the matrix elements of $u$: $1-\delta$ and $(2-\delta)/\sqrt{2}$, are also plotted.
}
\label{fig:uv}
\end{figure}

For later purposes, it is useful to show the operators constructed by the $t_{2g}$  occupation number 
$n_{xy},\ n_{yz}$, and $n_{zx}$; $n\equiv n_{xy}+n_{yz}+n_{zx}$, $u\equiv 2n_{xy}-n_{yz}-n_{zx}$, and $v\equiv 
\sqrt{3}(n_{yz}-n_{zx})$.

\begin{align}
	n &\simeq {\rm diag}(1,1,1-\delta,1-\delta,1-\delta,1-\delta),\\
	u &\simeq \small\begin{bmatrix}
		0 & 0 & \frac{2-\delta}{\sqrt{2}} & 0 & 0 & 0\\
		 & 0 & 0 & -\frac{2-\delta}{\sqrt{2}} & 0 & 0\\		
		 &  & 1-\delta & 0 & 0 & 0\\		
		 &  &  & 1-\delta & 0 & 0\\		
		 &  &  &  & -1+\delta & 0\\		
		 &  &  &  &  & -1+\delta\\				
	\end{bmatrix}, \label{eq:matu}\\
	v &\simeq \small\begin{bmatrix}
		0 & 0 & 0 & 0 & \frac{2-\delta}{\sqrt{2}} & 0\\
		 & 0 & 0 & 0 & 0 & \frac{2-\delta}{\sqrt{2}}\\		
		 &  & 0 & 0 & -1+\delta & 0\\		
		 &  &  & 0 & 0 & 1-\delta\\		
		 &  &  &  & 0 & 0\\		
		 &  &  &  &  & 0\\				
	\end{bmatrix}, \label{eq:matv}
\end{align}
where $\delta=\frac{3}{50}\varepsilon^2+O(\varepsilon^3)$, and as similarly to Eq.~(\ref{eq:matJz}), we have omitted the lower triangle part. 
For $\delta\to 0$, $n\to \mathbbm 1$ (an identity matrix) and $e_g$ quadrupole moments $(u,v)$ are reduced to the quadrupole moments $(q_u,q_v)$ constructed by the angular momentum ${\bm L}$ as 
\begin{align}
q_u=\frac{2L_z^2-L_x^2-L_y^2}{3},\quad 
q_v= \frac{\sqrt{3}(L_x^2-L_y^2)	}{3}.
\end{align}
Note that the $u$ and $v$ include the virtual processes accross the CEF excited quartet states, which is different from that calculated by the $L_{x,y,z}$ restricted within the $t_{2g}$ states. 
It is natural to obtain these expressions in terms of $\bm L$ for $\delta\to 0$ instead of those of the total angular momentum $\bm J$ since $n_{xy,yz,zx}$ does not depend on the spin. 
We note that $(u,v)$ possess the offdiagonal matrix elements $\pm \frac{2-\delta}{\sqrt{2}}$ between the ground states and the excited states and their magnitudes are larger than those among the ground states. This is the source of the anisotropy in the quadrupole moments and enables one to obtain ferro quadrupole moments under antiferro quadrupole orders. In Fig.~\ref{fig:uv}, the full expressions, i.e., without assumption of the small $\varepsilon(=\lambda/B_4)$ are shown as a function of $\varepsilon$. The offdiagonal elements for $u$ and $q_u$ are exactly the same, while the diagonal ones differ as $\varepsilon$ increases. 
For the realistic parameter  regime $\varepsilon\lesssim 0.5$, the difference between the full expression and the approximated one shown in Eqs.~(\ref{eq:matu}) and (\ref{eq:matv}) are quantitatively the same.  

The finite offdiagonal elements $\pm (2-\delta)/\sqrt{2}\sim \pm \sqrt{2}$ for $\varepsilon\lesssim 0.5$ leads to a finite cubic anisotropy in the local quadrupole free energy, which can be calculated by the local CEF 
model {via the Legendre transformation\cite{Ishitobi2023}} as 
\begin{align}
	F_q^{\rm loc}\sim F_{q0}^{\rm loc}+a (\phi_u^2+\phi_v^2) - b (\phi_u^3-3\phi_u\phi_v^2) + \cdots,
\end{align}
where $F_{q0}^{\rm loc}, a$, and $b$ are constants and $\phi_{u}$ and $\phi_v$ correspond to the quadrupole fields for $u$ and $v$, respectively. The coefficient of the cubic anisotropy $b$ is given by 
\begin{align}
	b\simeq \frac{1}{{2}\beta^2 (\frac{3}{2}\lambda)}  \left(\frac{2-\delta}{\sqrt{2}}\right)^2.
\end{align}
Here, $\beta$ is the inverse of temperature $T$. 
{This is the consequence of standard Landau expansion of quadrupole free energy for $\beta\lambda\gg 1$. The $T^2$ dependence of $b$ is meaningful near the quadrupolar transition temperature $T_q\simeq 33$ K and $T$ is usually replaced by $T_q$ for its phenomenological analysis.  We note} that $b$ is proportional to the square of the offdiagonal element in $u$ and $v$ in Eqs.~(\ref{eq:matu}) and (\ref{eq:matv}). The denominator $\frac{3}{2}\lambda$ represents that the perturbative processes to the CEF excited  doublet are important. {As for the opposite limit $\beta\lambda\sim 0$, $b\propto T$.} 


\subsection{Interactions in $t_{2g}$ manifolds}
In this subsection, we introduce the exchange interactions between the $t_{2g}$ electrons. The model used in this paper is basically given in Ref.~\onlinecite{Chen2010}, but for making this paper self-contained we summarize the model in the following. 

First, we discuss exchange interactions in the quadrupole sector. We use the quadrupole-quadrupole interactions introduced in  Ref.~\onlinecite{Chen2010} with slightly different notation\cite{Tsunetsugu2021,Hattori2022-ss} including two parameters $g_{\rm iso}$ and $g_{\rm ani}$ as 
\begin{align}
	H_q=\sum_{\langle i,j\rangle} (g_{\rm iso} \QQ_i\cdot \QQ_j+g_{\rm ani} \QQ_i\cdot {\mathsf K}_{ij}\QQ_j).
	\label{eq:Hq}
\end{align}
Here, ${\QQ}_i\equiv (u_i,v_i)$ is the $e_g$ quadrupole vector at the $i$th site and the sum runs over the nearest-neighbor sites on the fcc lattice. The bond directional anisotropic interactions are parameterized by the matrix ${\mathsf K}_{ij}$:
\begin{align}
	{\mathsf K}_{ij}&=\begin{bmatrix}
		-1 & 0\\
		0 & 1
	\end{bmatrix}&\equiv{\mathsf K}^{3}&\quad ij \ \text{bond} \parallel \text{$xy$ plane},\\
	{\mathsf K}_{ij}&=\begin{bmatrix}
		-c & s\\
		s & c
	\end{bmatrix}&\equiv{\mathsf K}^{1}&\quad ij \ \text{bond} \parallel \text{$yz$ plane},\\	
	{\mathsf K}_{ij}&=\begin{bmatrix}
		-c & -s\\
		-s & c
	\end{bmatrix}&\equiv {\mathsf K}^{2}&\quad ij \ \text{bond} \parallel \text{$zx$ plane},
\end{align} 
with $c\equiv \cos\frac{2\pi}{3}$ and $s\equiv \sin\frac{2\pi}{3}$. One can also rewrite them as 
\begin{equation}
	{\mathsf K}^{n}=-\hat{t}_{n}\hat{t}_{n}^{\ \rm T}, \quad \hat{t}_{n}^{\ \rm T} \equiv (\cos n\omega,\sin n\omega),\quad n=1,2, 3,
\end{equation}
with $\omega\equiv 2\pi/3$. Note that the unit vector $\hat{t}_{1,2,3}$ represents the projection operator to $3x^2-r^2$, $3y^2-r^2$, and $3z^2-r^2$ type orbital, respectively. 
The parameter $V$ in Ref.~\onlinecite{Chen2010} corresponds to 
$g_{\rm iso}=-7V/72$ and $g_{\rm ani}=-25V/72$. Since the ratio $g_{\rm ani}/g_{\rm iso}=25/7>2$, this naively suggests an antiferro ``$v$'' order (O$_{22}$ type) with the ordering vector at the X point: $(0,0,2\pi)$ from the results for $E_g$ non-Kramers doublet systems on the fcc lattice.\cite{Tsunetsugu2021,Hattori2022-ss} For a more general situation, the ratio $g_{\rm ani}/g_{\rm iso} \ne 25/7$, but in this paper we restrict ourselves to the case derived in Ref.~\onlinecite{Chen2010} since the modification in $g_{\rm ani}/g_{\rm iso}$ leads to just a qualitative difference for our purpose in this paper.

For the spin part, by using the spin operator for the $\rho=xy,yz$, and $zx$ orbital: $S_{i,\rho}^\mu\equiv \frac{1}{2}\sum_{\sigma\sigma'}d^\dagger_{i,\rho\sigma}(\hat{\sigma}^\mu)_{\sigma\sigma'} d_{i,\rho\sigma'}$, antiferromagnetic interactions are given as
\begin{align}
	H_{s1} =& J\sum_{\langle i,j\rangle}^{ xy\text{-plane}}\left(\sum_\mu S_{i,xy}^\mu S_{j,xy}^\mu -\frac{1}{4}n_{i,xy}n_{j,xy}\right)\nonumber\\
	& +\left( xy\ \to\ yz\ {\rm and}\ zx  \right).
	\label{eq:Hs1}
\end{align}
Here, $n_{i,\rho}\equiv \sum_\sigma d^\dagger_{i,\rho\sigma}d_{i,\rho\sigma}$ is the number operator for $\rho=xy,yz$, and $zx$ orbitals. There are also ferromagnetic interaction,
\begin{align}
	H_{s2} =& -J'\!\!\!\sum_{\langle i,j\rangle}^{ xy\text{-plane}}\left[ \sum_\mu S_{i,xy}^\mu (S_{j,yz}^\mu+S_{j,zx}^\mu)+i\leftrightarrow j\right]\nonumber\\
	& +\frac{3J'}{2}\sum_{\langle i,j\rangle\in xy
	\text{-plane}}n_{i,xy}n_{j,xy}\nonumber\\
	& +\left( xy\ \to\ yz\ {\rm and}\ zx  \right).	\label{eq:Hs2r}
\end{align}
$n_{i\rho}$ is represented by the quadrupole operators as
\begin{align}
	n_{i,xy} &= \frac{1}{3}(n_i+u_i),\\
	n_{i,yz} &= \frac{1}{3}\left(n_i-\frac{1}{2}u_i+\frac{\sqrt{3}}{2}v_i\right),\\
	n_{i,zx} &= \frac{1}{3}\left(n_i-\frac{1}{2}u_i-\frac{\sqrt{3}}{2}v_i\right),	\label{eq:Hs2}
\end{align}

The terms consisting of $n_{i,\rho}$'s in Eqs.~(\ref{eq:Hs1}) and {(\ref{eq:Hs2r})} can be rewritten by the quadrupole forms as 
\begin{align}
	H'_{q}=-\frac{J-6J'}{72} \sum_{\langle i,j\rangle}\QQ_i \cdot (\mathbbm 1-\mathsf{K}_{ij}) \QQ_j, \label{eq:Hq2}
\end{align}
which renormalizes $g_{\rm iso}$ and $g_{\rm ani}$ in Eq.~(\ref{eq:Hq}).

Summing up Eqs.~(\ref{eq:Hq}), (\ref{eq:Hs1}), and {(\ref{eq:Hs2r})} with Eq.~(\ref{eq:Hq2}), we obtain the total nearest-neighbor exchange Hamiltonian as
\begin{align}
	H_{\rm int}=&\sum_{\langle i,j\rangle} \left(\tilde{g}_{\rm iso} \QQ_i\cdot \QQ_j+\tilde{g}_{\rm ani} \QQ_i\cdot {\mathsf K}_{ij}\QQ_j \right)\nonumber\\
	&+(J+2J')\sum_{\rho}\sum_{\langle i,j\rangle}^{\rho{\text{-plane}}}
	\bm{S}_{i,\rho}\cdot \bm{S}_{j,\rho}\nonumber\\
	&-J'\sum_{\rho}\sum_{\langle i,j\rangle}^{\rho\text{-plane}}
	(\bm{S}_{i,\rho}\cdot \bm{S}_{j}+\bm{S}_{j,\rho}\cdot \bm{S}_{i}),
	\label{eq:Hint}
\end{align}
where $\rho=xy,yz$, and $zx$. We have introduced the $t_{2g}$  spin operators at the $i$ site: $\bm{S}_{i}=\sum_\rho \bm{S}_{i,\rho}$ and the renormalized quadrupole interactions $\tilde{g}_{\rm iso}=g_{\rm iso}-\frac{1}{72}(J-6J')$ and 
$\tilde{g}_{\rm ani}=g_{\rm ani}+\frac{1}{72}(J-6J')$.

\section{Mean-field results}\label{sec:results}
In this section, we show the results of the mean-field analysis of the model (\ref{eq:Hint}) with the local part of the Hamiltonian (\ref{eq:Hloc}). In Sec.~\ref{sec:Hmag0}, we demonstrate the results for the magnetic field $\bm{h}={\bf 0}$. For ${\bm h}={\bf 0}$, the CEF excited quartet plays 
just a minor role as is evident from the small factors 
$\delta\simeq \frac{3}{50}\varepsilon^2$ and $\alpha\simeq \frac{3\sqrt{2}}{100}\varepsilon^2$ in the matrix elements in the Hamiltonian (\ref{eq:Hint}) in addition to the energy gap $D\sim 5B_4$. See Eqs.~(\ref{eq:matu}), (\ref{eq:matv}), and (\ref{eq:matS1})--(\ref{eq:matS3}). 
Thus, one can consider 
 the $\varepsilon=\lambda/B_4\to 0$ limit neglecting the CEF excited quartet states with setting $\epsilon_{3/2}=0$, $\epsilon_{1/2}=\frac{3\lambda}{2}$, and $\epsilon_{3/2}^\prime\simeq 5B_4\to \infty$ in Eqs.~(\ref{eq:e12}), (\ref{eq:e32a}), and (\ref{eq:e32b}), respectively.  Here, we shift the energy in order to set the energy of the CEF ground quartet states to 0. 
The matrix elements of various operators are also simplified in this limit: These simplifications are justified for ${\bm h}={\bf 0}$. In Sec.~\ref{sec:Hmag}, we will discuss the properties under finite magnetic fields $\bm{h}\ne {\bf 0}$. It turns out that one needs $O(\varepsilon)$ terms in the Zeeman energy. 

\subsection{Two-sublattice mean-field approximation for zero magnetic field}\label{sec:Hmag0}
Throughout this paper we assume two-sublattice orders, which are indeed observed experimentally in the double-perovskite compounds. {This is because our primary purposes in this paper are to demonstrate the electronic mechanism for the induced ferro quadrupole moments under antiferro quadrupole orders (Sec.~\ref{sec:Hmag0}) and the finite magnetic moments in $J_{\rm eff}=3/2$ systems (Sec.~\ref{sec:Hmag}). }
In recent analysis, Ref.~\onlinecite{Svoboda2021-zu} have carried out similar calculations with a four-sublattice unit cell. However, there is no comment about the finite ferro quadrupole moments in the antiferroic quadrupole phase, although the reason is unclear. For our purposes, it is sufficient to use the two-sublattice unit cell. {In some parameter sets used in the following sections, the four-sublattice antiferro magnetic orders\cite{Svoboda2021-zu}, which correspond to double-$\bq$ magnetic orders with induced a single-$\bq$ quadrupole moments, are more stabilized than FM110 states (see below), when one applies the 4-sublattice approximation. In this sense, our results within the two-sublattice orders are regarded as simple extension with the excited states from  the $J_{\rm eff}=3/2$ model in Ref.~\onlinecite{Chen2010}. 
Nevertheless, the comparison between the results within the two-sublattice approximation and the experiments for the FM110 orders in Sec.~\ref{sec:Hmag} possesses important information since regardless of which types of the microscopic interactions favor the experimentally observed FM110 phase, qualitative properties in the FM110 phase are unchanged. The detail analysis about the four-sublattice orders is one of our future problems.}

Let us take the two sublattice position in the unit cell as A: $(0,0,0)$ and B: $(0,\frac{1}{2},\frac{1}{2})$ and assume the ordered configurations are uniform on the $xy$ planes: $(x,y,n)$ and $(x+\frac{1}{2},y,n+\frac{1}{2})$, where the lattice constant is set to unity and $n$ is an integer and $x$ and $y$ represent the site positions on the $z=n$ plane. This choice corresponds to the domain with its ordering wavevector ${\bm k}^*_3\equiv (0,0,2\pi)$. The expressions for the mean-field Hamiltonian are trivially obtained and listed in Appendix~\ref{app:Hmf}. 
The inclusion of the CEF excited doublet in the model Hamiltonian causes only small effects on the overall feature 
of the phase diagram and the magnitude of the primary order parameters. Thus, 
the qualitative results here are almost identical to those in Ref.~\onlinecite{Chen2010} except for the ferroic quadrupole moments.

\begin{figure}[t!]
\begin{center}
\includegraphics[width=0.45\textwidth]{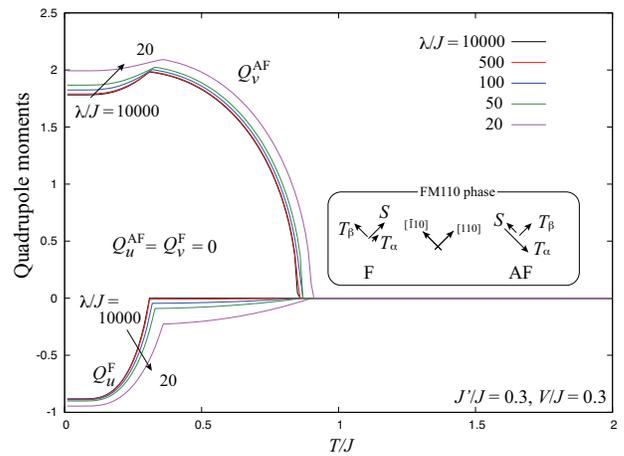}
\end{center}
\caption{Temperature dependence of the quadrupole moments $Q_u^{\rm F}$ and $Q_v^{\rm AF}$ for $J'/J=0.3,\ V/J=0.3$, and $\lambda/J=10^4,\ 500,\ 100,\ 50$, and $20$. $Q_v^{\rm F}=Q_{u}^{\rm AF}=0$ and are not shown.
}
\label{fig:QvsT1}
\end{figure}

Figure 2 shows the temperature dependence of the ferro and antiferro quadrupole moments defined as
\begin{align}
	{\bm Q}^{\rm F}\equiv 	\langle {\bm Q} \rangle_{\rm A}+\langle {\bm Q} \rangle_{\rm B},\quad 
    {\bm Q}^{\rm AF}\equiv 	\langle {\bm Q} \rangle_{\rm A}-\langle {\bm Q} \rangle_{\rm B}.\label{eq:F-AF}
\end{align}
Here $\langle \cdot \rangle_{\rm A(B)} $ indicates the expectation value calculated by the A(B)-site local mean-field Hamiltonian. 
The data for $\lambda/J=10^4\gg 1$ corresponds to those shown in Fig.~8 in Ref.~\onlinecite{Chen2010}. For these parameter sets, the system undergoes a phase transition into a pure quadrupolar phase with the primary antiferroic and induced ferroic ones (AFQ+$f$q) at $T/J\simeq 0.9$ and another transition at $T/J\simeq 0.3$ into a magnetic phase (FM110). The order parameter configuration of the FM110 phase is schematically shown in Fig.~\ref{fig:QvsT1}: the $x$ and $y$ components of the ferro/antiferro dipole ${\bm S}^{\rm F,AF}$ and ferro/antiferro octupole ${\bm T}_{\alpha,\beta}^{\rm F,AF}$ moments take finite values. The definition of these multipole moments are given in Appendix~\ref{app:multipole} and we use the definition of ``F/AF'' similarly to Eq.~(\ref{eq:F-AF}). Note that the realistic value of $\lambda$ in Fig.~\ref{fig:QvsT1} is e.g., $\lambda/J=100$ baring the magnetic transition temperature $\sim$20 K in mind. One can see a finite $Q_{u}^{\rm F}\lesssim -0.1$ inside the AFQ+$f$q phase for $0.3\lesssim T/J \lesssim 0.9$. It is quite natural that as $\lambda$ decreases and thus the energy of the excited doublet $\epsilon_{1/2}$ lowers, the ferro moment $|Q_u^{\rm F}|$ increases. It is interesting that the effect of the excited state is not negligible even if the excited state energy is 100 times larger than the transition temperature into the quadrupole order. The presence of the excited doublet also affects the transition temperatures and the two transition temperatures slightly increase upon decreasing $\lambda$ but these are minor points.

\begin{figure}[t!]
\begin{center}
\includegraphics[width=0.5\textwidth]{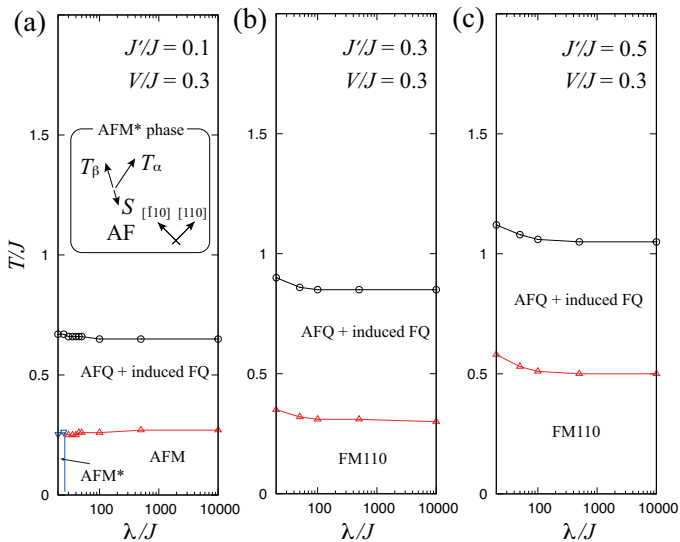}
\end{center}
\caption{$T$-$\lambda$ phase diagram for $V/J=0.3$, and (a) $J'/J=0.1$, 
(b) $0.3$, and (c) 0.5. Inset in (a): the antiferromagnetic configurations 
${\bm S}^{\rm AF}$, ${\bm T}_{\alpha}^{\rm AF}$, and ${\bm T}_{\beta}^{\rm AF}$ in the AFM* phase.}
\label{fig:phaseDiagram1}
\end{figure}

Figure \ref{fig:phaseDiagram1} shows $T$-$\lambda$ phase diagram for $V/J=0.3$ and $J'/J=0.1$--$0.5$. For these parameter sets, the AFQ+$f$q phase robustly appears from the normal (paramagnetic) state via the second-order transition. For $J'/J=0.3$ and $0.5$, the ground state is the FM110, whose configuration is shown in the inset of Fig.~\ref{fig:QvsT1}. For $J'/J=0.1$, the ground states change with varying $\lambda$. When $\lambda$ is large, the planer antiferromagnetic (AFM) order takes place, where  $S^{\rm AF}_y/S^{\rm AF}_x=T_{\alpha,y}^{\rm AF}/T_{\alpha,x}^{\rm AF}=-T_{\beta,y}^{\rm AF}/T_{\beta,x}^{\rm AF}$ and the other multipole moments are zero. As noted in Ref.~\onlinecite{Chen2010}, owing to an accidental degeneracy there is no anisotropy and $S^{\rm AF}_{x,y}=0$ at $T=0$, while the $x$ and $y$ components of ${\bm T}^{\rm AF}_{\alpha,\beta}$ are finite. 
{This feature continues even for finite $\lambda$. This is because the ground-state wavefunction at each sublattice (A and B) in the AFM phase for finite $\lambda$ is exactly the same as that at $\lambda=\infty$, which is the eigenstate of the quadrupole moment $\langle \QQ\rangle_{\rm A,B}=(-1+\delta,0)$. As is evident from  Eq.~(\ref{eq:matu}), the matrix $u$ does not have finite matrix elements between the excited states and the fifth and sixth states, while $v$ in Eq.~(\ref{eq:matv}) does. In the language of ${\bm S}_{\rho}$, the ground state is characterized by 
$\langle{\bm S}_{xy}\rangle_{\rm A,B}={\bf 0}$, $\langle S_{yz,zx}^z\rangle_{\rm A,B}=0$, $\langle S_{yz}^{\mu}\rangle_{\rm A}=-\langle S_{yz}^\mu\rangle_{\rm B}=-\langle S_{zx}^\mu\rangle_{\rm A}=\langle S_{zx}^\mu\rangle_{\rm B}$ with $\mu=x$ or $y$. They ensure that the state with $\langle \QQ\rangle_{\rm A,B}=(-1+\delta,0)$ remains to be decoupled from the excited states since there is only a term proportional to $S_{i,yz}^\mu-S_{i,zx}^{\mu} (\mu=x,y)$ in the magnetic sector of the mean-field Hamiltonian at the site $i$. See Appendix \ref{app:Hmf}. This has indeed no matrix element between the states with $u=-1+\delta$ and the excited states as in $u$. See their expressions in Eqs.~(\ref{eq:matSyz1}), (\ref{eq:matSyz2}), (\ref{eq:matSzx1}), and (\ref{eq:matSzx2}).
}

In contrast, for small $\lambda$, the low-$T$ phase is another antiferromagnetic phase (AFM*) as schematically shown in the inset of Fig.~\ref{fig:phaseDiagram1}(a). The order parameters for the AFM* phase are similar to those in the AFM phase, but there are finite $S_{x,y}^{\rm AF}$ moments even at $T=0$. To realize the AFM* phase, the spin-orbit coupling $\lambda$ is tuned to be {$\lambda\lesssim 30 J$, where a finite quadrupole moments $v$ appears and thus the symmetry is lowered. This also causes a finite mixing between the states with $u=-1+\delta$ and the excited states. Thus, the ground-state wavefunction in the AFM state is no longer eigen state of the mean-field Hamiltonian. Then, a different state becomes the ground state with a finite ${\bm S}^{\rm AF}$. Indeed, such transition has been studied in the non-Kramers $\Gamma_3$ system in our previous study\cite{Hattori2022-ss}, where a kind of topological protection is important. However, the value of the spin-orbit coupling is not realistic, and thus,} we do not analyze it in detail here. 

Let us now examine the $T$-$V$ phase diagram for the realistic value of $\lambda \simeq 100J$.  
There are three phases: the AFM, the AFQ+$f$q, and the FM110 phases in Fig.~\ref{fig:phaseDiagram2}. For visualizing the magnitude of the $Q^{\rm F}_u$ in the ordered phases, $\sqrt{-Q^{\rm F}_u}$ is also depicted in Fig.~\ref{fig:phaseDiagram2} as colormap. The reason for using the ``square-root" is just due to the technical one to show the finiteness in the AFQ+$f$q  phase, where the $|Q_{u}^{\rm F}|$ is much smaller than that in the other ordered phases. We have checked that even for $\lambda/J=10000$ ignoring the excited states, the phase diagram is semiquantitatively the same as that shown in Fig.~{\ref{fig:phaseDiagram2}. The difference is that the $Q_u^{\rm F}\simeq 0$ in the AFQ+$f$q for $\lambda/J=10000$, which corresponds to the results in Ref.~\onlinecite{Chen2010}.

\begin{figure}[t!]
\begin{center}
\includegraphics[width=0.45\textwidth]{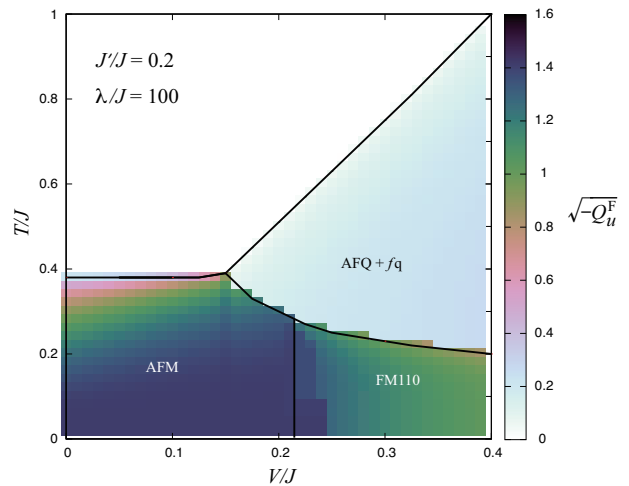}
\end{center}
\caption{$T$-$V$ phase diagram for $J'/J=0.2$. The lines represent the phase boundaries. The color map shows $\sqrt{-Q_u^{\rm F}}$, where $Q_u^{\rm F}\le 0$ for any $V$ and $T$. The square-root of $-Q_u^{\rm F}$ is used just for visualization of the small value of $Q_u^{\rm F}$ in the AFQ phase.}
\label{fig:phaseDiagram2}
\end{figure}

\subsection{Finite magnetic fields}\label{sec:Hmag}
Under the magnetic field $\bm h=(h_x,h_y,h_z)$, which includes the Bohr magneton $\mu_{\rm B}$ and is related to the real magnetic field $\bm{H}$ as $\bm{h}=\mu_{\rm B}\bm{H}$, we introduce the Zeeman coupling with the electronic $g$ factor with $g=2$ as 
\begin{align}
		H_{\rm Z} =-\sum_{\sigma,\sigma',m,m'} {\bm h} \cdot ({\bm M})_{\sigma\sigma',mm'}d_{m\sigma}^\dag d_{m'\sigma'}. \label{eq:Zeeman}
\end{align}
Here, the magnetic moment operator is defined as $({\bm M})_{\sigma\sigma',mm'}\equiv [(\mathbbm{1})_{\sigma\sigma'}({\bm L})_{mm'}+2\bm{s}_{\sigma\sigma'}(\mathbbm{1})_{mm'}]$. In actual calculations for the six states manifold, $\bm{M}$ reduces to the matrix form listed in Eqs.~(\ref{eq:matS1})--(\ref{eq:matS3}) and Eqs.~(\ref{eq:matL1})--(\ref{eq:matL3}).

To determine which ordering wavevectors are stabilized under the magnetic fields within the two-sublattice mean-field approximation, 
we repeat the calculations with different ordering wavevector set up: 
${\bm k}^*_1=(2\pi,0,0)$, ${\bm k}^*_2=(0,2\pi,0)$, and ${\bm k}^*_3=(0,0,2\pi)$ and compare their free energies. We note that the diagonal matrix elements of ${\bm M}$ in the bases of the eigenstates of the Hamiltonian (\ref{eq:Hloc}) are $O(\varepsilon)$ in the CEF ground-state quartet. For example, the diagonal elements of $M_z$ is diag$(1,-1,\frac{4}{5}\varepsilon+\frac{1}{50}\varepsilon^2,-\frac{4}{5}\varepsilon-\frac{1}{50}\varepsilon^2,0,0)+O(\varepsilon^3)$. Thus, the half of the quartet state with $u\simeq 1$ has a finite dipole moment $\sim \frac{4}{5}\varepsilon \mu_{\rm B}\simeq 0.4\mu_{\rm B}$, which has not been recognized well so far.

Before discussing the numerical data, we should note the followings. 
Using the $O(\varepsilon)$ terms in the $\bm{M}$ means that we treat the CEF ground state quartet as {\it effective} $t_{2g}$ state with hybridizing to the excited quartet. This treatment is valid if the interaction (\ref{eq:Hint}) are not modified significantly when the processes related to the $e_g$ electrons are also taken account. In this study, we do not estimate such processes and we keep the interaction form derived within the $t_{2g}$ states as a starting approximation. The more sophisticated analysis remains as one of our future problems.

Figure \ref{fig:mag} shows the magnetization $M_h\equiv \langle \bm{M} \rangle \cdot \bm{h}/h$ and $|\langle \bm{M} \rangle|$ 
for $J'/J=0.3$ and $V/J=0.27$  
as a function of $h=|\bm{h}|$ for three high-symmetric directions 
$\bm{h}\parallel [001],\ [110]$, and $[111]$. At $h=0$, the ground state is the FM110 phase 
and the intermediate-temperature phase is the 
AFQ+$f$q phase for this parameter set. 
As for the scale of the magnetic field, 
$h/J=0.1$ corresponds to $\sim 10$ T if we set 
the magnetic transition temperature 
$T_m\sim 0.34J\sim 20$ K. There are three 
equivalent domains for the FM110-type phases: 
FM110, FM101, and FM011. As can be 
trivially understood, these phases 
have its magnetization 
along [110], [101], and [011] directions, 
respectively. A finite magnetic field selects 
some of the three domains and the results are 
shown only for those with the lowest free energy. 

 Owing to the difference in the field direction 
 and that for the spontaneous one for 
 $\bm{h}\parallel [001]$ and [111], 
 $M\ne M_h$, while $M= M_h$ for  
 $\bm{h}\parallel [110]$. One can notice that 
 the increase in $M_h$ as increasing $h$ is 
 steepest for $\bm{h}\parallel [001]$. This is
  related to the fact that the ferro quadrupole 
  moment $Q_u^{\rm F}$ is large in the FM110 
  phase (Fig.~\ref{fig:QvsT1}) and the increase 
  in $M_h=M_z$ is more favored than that 
  for $\bm{h}\parallel [111]$. 
 In the AFQ+$f$q phase, the magnetic field 
 also selects the two of the domains: $\bm{k}_{1}$ 
 or $\bm{k}_{2}$.
 However, the energy difference between the domains with $\bm{k}_{1,2}$ and $\bm{k}_3$ is so tiny and 
 it is not physically important in actual 
 situations, where there are many aspects 
 such as electron-lattice couplings.


\begin{figure}[t!]
\begin{center}
\includegraphics[width=0.45\textwidth]{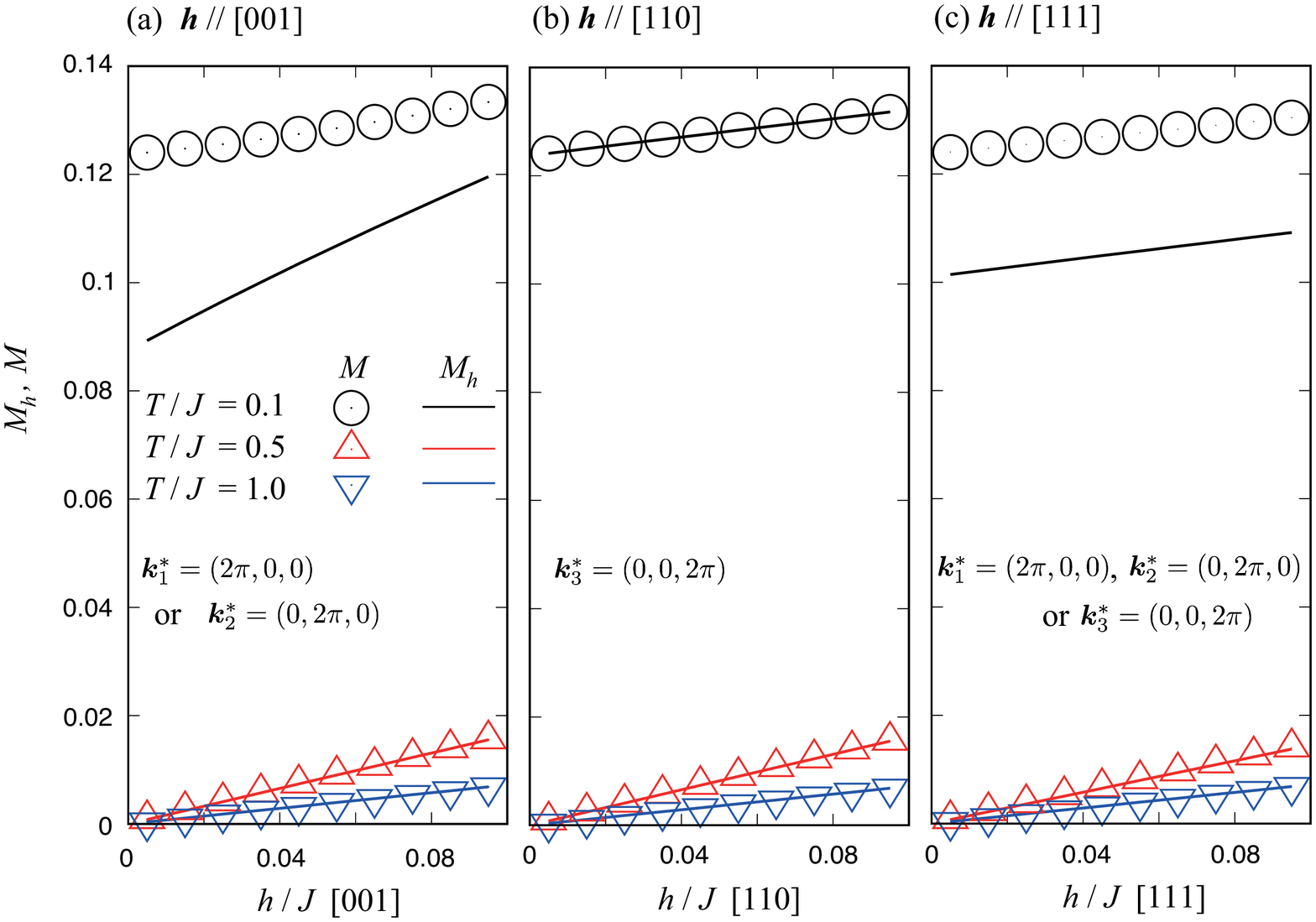}
\end{center}
\caption{Magnetization curve for $J'/J=0.3$ and $V/J=0.27$. (a) ${\bm h}\parallel [001]$, (b) ${\bm h}\parallel [110]$, and (c) ${\bm h}\parallel [111]$. The magnetization $M_h$ is defined as $M_h\equiv \langle \bm{M}\rangle\cdot \bm{h}/h$, with $h\equiv |\bm{h}|$ and plotted by lines, while the absolute value $M\equiv |\langle\bm{M}\rangle|$ is indicated by symbols. The domain with the lowest free energy is considered; (a) ${\bm k}_{1}$ or ${\bm k}_{2}$ domains, (b) ${\bm k}_{3}$ domain, and (c) ${\bm k}_{1}$, ${\bm k}_{2}$, or ${\bm k}_{3}$ domains. The temperatures shown correspond to the FM110 phase ($T/J=0.1<T_m/J\simeq 0.34$), the AFQ+$f$q phase ($T/J=0.5$), and the normal phase ($T/J=1.0$).}
\label{fig:mag}
\end{figure}


\section{Discussions}\label{sec:dis}
In the following subsections, we discuss two aspects. The first is the induced ferro quadrupole moments under the antiferro quadrupole orders. The second is the magnitude of the magnetic moment, which vanishes in the model of $J_{\rm eff}=3/2$. 

\subsection{Ferro quadrupole moments}
Let us compare the results in Sec.~\ref{sec:results} and the experimental data in the double perovskite compound focusing on the ferro quadrupole moments observed in 
Ba$_2$MgReO$_6$\cite{Hirai2020}.
The relative magnitude of the antiferro quadrupole $Q^{\rm AF}_v$ and the ferro quadrupole $Q^{\rm F}_u$ moments can be indirectly estimated by observing O displacement $(\varepsilon_u,\varepsilon_v) $ corresponding to $(u,v)$ in the ordered phases. At $T=6$ K, i.e., inside the magnetically-ordered phase, Hirai et al., reported that there is about 0.4 \% elongation of the oxygen octahedron along the $z(c)$ axis in average. Here ``average'' means that the analysis without inplane $v$ type displacement. With the further analysis including this inplane displacements, 
 they estimated that the ratio 
 $v/u \sim Q^{\rm AF}_v/Q^{\rm F}_u
 \sim 4$. Here, we have assumed a linear 
 relation between the oxygen displacement 
 and the $d$-electron quadrupole 
 moments, ignoring anharmonic couplings. 
 In the Supplimental Material in 
 Ref.~\onlinecite{Hirai2020}, 
 the data of average oxygen positions at 
 $T=25$ K inside the quadrupolar phase 
 are also listed. One 
 can estimates 0.18 \% elongation of the 
 oxygen octahedron along the $z(c)$ axis 
 in average. This naively leads to the 
 ratio  $Q^{\rm AF}_v/Q^{\rm F}_u\sim 8$. 
 Although there is ambiguity about the 
 quadrupole-displacement coupling, which 
 can be anisotropic even in the first-order 
 in $Q_{u,v}$, the finite values 
 of $Q_u^{\rm F}$ shown in 
 Fig.~\ref{fig:QvsT1} are qualitatively consistent with this.  
 Indeed, when setting $J\sim 50$ K 
 and $\lambda${$/J$}$=50$--$100$, 
 the AFQ order appears at $T_q\sim 40$ K 
 and the magnetic one at $T_m\sim 25$ K in 
 Fig.~\ref{fig:QvsT1}. 
 In actual situation, one should 
 take into accoount couplings between 
 the electrons and the oxygen displacement 
 as discussed recently\cite{Iwahara2022-mt,Iwahara2018-mt}. 
 Nevertheless, this result demonstrates 
 that the contribution of the excited states 
 on the induced ferro quadrupole moments 
 in the AFQ state is a noticeable  
 in the double perovskite materials.

 \subsection{Effective magnetic moment}
We now discuss the magnetic properties of this system. It is well known that the magnetic moment $\bm{M}$ in the effective $J_{\rm eff}=3/2$ theory, i.e., $\varepsilon\to 0$ limit, vanishes. This is one of the weak points in the theory 
and some discussions about the impact of the orbital orders on the magnetic susceptibility have been done recently\cite{Svoboda2021-zu}. 
Let us estimate the effective moment $\mu_{\rm eff}$ defined by the high-temperature asymptotic form of the magnetic susceptibility as 
\begin{eqnarray}
\mu_{\rm B}M_z&\simeq &\mu_{\rm B}\frac{\frac{4\varepsilon  }{5}(e^{\frac{4}{5}\varepsilon h/T}-e^{-\frac{4}{5}\varepsilon h/T})}{e^{\frac{4}{5}\varepsilon  h/T}+e^{-\frac{4}{5}\varepsilon h/T}+2}
\simeq \frac{(\frac{\sqrt{3}}{\sqrt{2}}\frac{4}{5}\varepsilon \mu_{\rm B})^2}{3 T} \frac{h}{\mu_{\rm B}},\ \ \ \ \ \ 	\label{eq:Mu_eff}
\end{eqnarray}
where we have assumed $\bm{h} \parallel 
[001]$.  Thus, 
\begin{align}
\mu_{\rm eff}=\frac{4\sqrt{3}}{5\sqrt{2}}\varepsilon \mu_{\rm B}
=\frac{4\sqrt{2}}{\sqrt{3}}\left(\frac{\frac{3}{2}\lambda}{5B_4}\right)\mu_{\rm B}.	\label{eq:MuEff2}
\end{align}
As is evident from the factor $\frac{3}{2}\lambda/(5B_4)$, this arises from the hybridization between the ground and  excited quartets. Remember that the induced ferro quadrupole moments discussed in the previous subsection arises from the offdiagonal elements of the quadrupole operators between the ground state quartet and the first excited doublet. 
Substituting the values for Rb$_2$TaCl$_6$ with $D\simeq 5B_4 \simeq 3.2$ eV and $\lambda\simeq 0.27$ eV leading to $\varepsilon\simeq 0.42$, the effective moment is estimated as $\mu_{\rm B}\simeq 0.41\mu_{\rm B}$, which is similar to the observed one $0.27\mu_{\rm B}$\cite{Ishikawa2019-xg}.  
When using the typical values for the double-perovskites,
$3\lambda/2\sim 0.5$ eV \cite{Ahn2017-hi,Paramekanti2018-pb} and $D\sim 5$ eV\cite{Oh2018-sn},  one finds 
$\mu_{\rm eff}\simeq 0.49\mu_{\rm B}$. 
The present estimation is not far from the observed one  $\mu_{\rm eff}\simeq 0.68\mu_{\rm B}$\cite{Hirai2019-en} in Ba$_2$MgReO$_6$.
 The remaining discrepancy might be improved by taking into account the effects of ligand ions\cite{Ahn2017-hi} and/or the vibronic degrees of freedom\cite{Iwahara2022-mt}. The fluctuation effects neglected in the present analysis also affect the magnitude of the magnetic moment quantitatively. Analyses including these more sophisticated aspects are one of our future problems.  
 
Compared to the estimation of $\mu_{\rm eff}$, the results of the ordered magnetic moment shown in Fig.~\ref{fig:mag} is not satisfactory in the present mean-field analysis. As discussed in Fig.~\ref{fig:mag}, which roughly corresponds to the parameter set for Ba$_2$MgReO$_6$ with setting  $J\sim 50$ K, $M_h$ and $M$ are $\sim 0.1$. This is only $\sim 20$--30 \% of the observed moment in Ba$_2$MgReO$_6$\cite{Hirai2019-en}. As shown in Fig.~\ref{fig:QvsT1}, $Q_u^{\rm F}<0$ and this corresponds to larger occupation in the $yz$ and $zx$ orbitals at $h=0$. In terms of the four states in the ground state quartet, the energy of the $\Downarrow$ states ($J^z_{\rm eff}=\pm 3/2$) is lower than that for the $\Uparrow$ states ($J^z_{\rm eff}=\pm 1/2$). See Eqs. (\ref{eq:A3})--(\ref{eq:A6}) 
and the matrix $u$ in Eq.~(\ref{eq:matu}). The point is that these  $\Downarrow$ states have no moment even with the $O(\varepsilon^2)$ correction. Thus, the ordered moment is smaller compared with the effective moment even without the  additional $\sqrt{3}$ factor in the definition of $\mu_{\rm eff}$: $0.49/\sqrt{3}=0.28>0.1$. Recently, Zhang et al., have proposed that the quadrupole configurations in the AFQ+$f$q phase modifies the spin-spin exchange interactions and the Dzyaloshinskii-Moriya type interaction induced in the AFQ+$f$q phase stabilizes the magnetic order of FM110 and obtained a reasonable value of the ferro moments,\cite{Zhang2023-fq}  although the relation between the CEF excited quartet states are not clear.

In Ref.~\onlinecite{Ahn2017-hi}, the orbital angular momentum renormalization was discussed. Owing to the extended $d$ like orbital including the surrounding $p$ orbitals at the oxygen sites, ${\bm M}=2{\bm S}+{\bm L}\to 2{\bm S}+\gamma{\bm L}$ with $\gamma=0.536$ for Ba$_2$MgReO$_6$. Although it is unclear whether their analysis without the cubic anisotropy includes the effect of the local excited states shown in this paper, let us here qualitatively examine 
how this renormalization effect influences the results in our model. By introducing $\gamma$ in Eq.~(\ref{eq:Zeeman}), the value of $M$ and $M_h$ change as decreasing $\gamma$ from $\gamma=1$. For the parameters shown in Fig.~\ref{fig:mag}, firstly $M$ and $M_h$ decrease and vanish at $\gamma\sim 0.8$ and then increase. At $\gamma=0.5$, $M_h\sim 0.25$ as shown in Fig.~\ref{fig:mu_eff}. In actual situation in real materials, not only the Zeeman energy (\ref{eq:Zeeman}), but also the interactions (\ref{eq:Hint}) must be modified. Nevertheless, a finite $\varepsilon$ modifies the orbital 
character in the presence of the renormalization owing to $\gamma$. As shown in Fig.~\ref{fig:mu_eff}, the orbital contributing to 
the magnetic moment is $\sim \Uparrow$ 
state ($J^z_{\rm eff}=\pm 1/2$) for $\gamma\sim 1$, while $\Downarrow$ components  ($J_{\rm eff}^z=\pm 3/2$) increase as lowering $\gamma$. For the ferro quadrupolar phase observed in Rb$_2$TaCl$_6$ and Cs$_2$TaCl$_6$, $Q_u^{\rm F}>0$ and $\gamma\sim 0.8$ is estimated\cite{Ishikawa2019-xg}. Since $Q_u^{\rm F}>0$, $\Uparrow$ components (i.e., $J_{\rm eff}=\pm 1/2$) are energetically favored both by the uniform distortions and magnetic fields.

For smaller $\gamma$, the correction owing to the excited state proportional to $\varepsilon$ is minor in the present simplified analysis. However, for large $\varepsilon$, $\mu_{\rm eff}\sim 0.5\mu_{\rm B}$ without the correction owing to the ligand ions. It is expected that the combined corrections by the excited states and the ligand ions influence the orbital character of the magnetic moment in more complex ways than discussed here. The theory describing 
the magnetic moment in the $J_{\rm eff}=3/2$ model requires more elaborate treatments and the interpretation of the experimental results should also be reconsidered. 
In this respect, it is highly important to clarify the orbital profile, i.e., the ratio between $\Uparrow$ and $\Downarrow$, of the magnetic moments as analyzed in, e.g., LiV$_2$O$_4$\cite{Shimizu2012-nq}.


\begin{figure}[t!]
\begin{center}
\includegraphics[width=0.45\textwidth]{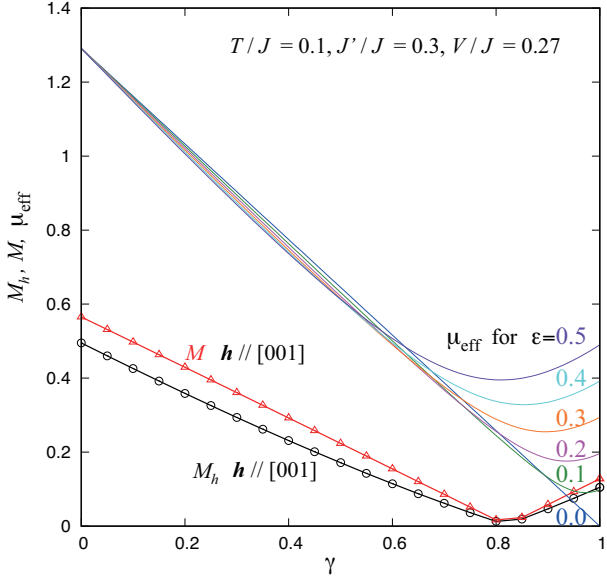}
\end{center}
\caption{$\gamma$ dependence of $M,\ M_h$ for $T/J=0.1$ K, $h/J=0.05$ along $\bm{h}\parallel [001]$, $J'/J=0.3$ and $V/J=0.27$. The effective moment $\mu_{\rm eff}$'s given by Eq.~(\ref{eq:MuEff2}) for several values of $\varepsilon$ are also drawn.}
\label{fig:mu_eff}
\end{figure}


\section{Summary}\label{sec:sum}
We have clarified that the ferroic components 
of $O_{20}$ are induced under the antiferroic 
$O_{22}$ orders in the fcc lattice model. 
We have microscopically 
demonstrated the mechanism of the induced ferroic moments.
This is in fact quite simple: just taking into account the CEF excited doublet state. 
It is important that the excited state 
located at $\sim 5000$ K, $\sim 100$ times 
larger scale than the transition temperature 
$T_q\sim 30$ K, can influence the order parameter 
via generating the local anisotropy. 
We have also clarified that the magnetic dipole moment 
in the $J_{\rm eff}=3/2$ model does not vanish in 
the realistic values of the spin-orbit coupling and crystalline electric 
field, when the CEF excited quartet states 
are taken into account. 
Since these aspects have not been seriously 
considered so far, we believe our results shed a 
renewed light on the orbital orders in 
related correlated systems. 

\section*{Acknowledgment}
This work was supported by JSPS KAKENHI 
(Grant No. JP21H01031 and No. JP23H04869).

\appendix

\section{Wavefunctions}\label{app:wavefunc}
We list the lowest 6 local eigenstates for $\lambda>0$ and $B_4>0$. They are mainly constructed by the $t_{2g}$ orbitals in the parameter regime relevent to our considerations. In terms of $m\sigma$ bases $\{2\uparrow,1\uparrow,0\uparrow,-1\uparrow,-2\uparrow,2\downarrow,1\downarrow,0\downarrow,-1\downarrow,-2\downarrow \}$, the local eigenstates for the $\Gamma_7$ with the energy $\epsilon_{1/2}$ and the $\Gamma_8$ with $\epsilon_{3/2}$ are expressed as 
\begin{align}
	|\Gamma_7 +\rangle:\ &\{0,-2,0,0,0,-1,0,0,0,1\},\\
	|\Gamma_7 -\rangle:\ &\left\{-1,0,0,0,1,0,0,0,2,0\right\},\\
	|\Gamma_8 +\Uparrow\rangle:\ &\left\{0,\frac{\varepsilon}{4}+\frac{w}{4}+\frac{1}{2},0,0,0,-\frac{\varepsilon}{2}-\frac{w}{2},0,0,0,1\right\},\label{eq:A3}\\
	|\Gamma_8 -\Uparrow\rangle:\ &\left\{1,0,0,0,-\frac{\varepsilon}{2}-\frac{w}{2},0,0,0,\frac{\varepsilon}{4}+\frac{w}{4}+\frac{1}{2},0\right\},\label{eq:A4}\\
	|\Gamma_8 -\Downarrow\rangle:\ &\left\{0,0,0,-1-\frac{\varepsilon}{10}-\frac{w}{2},0,0,0,\frac{\sqrt{6}}{5} \varepsilon,0,0\right\},\label{eq:A5}\\
	|\Gamma_8 +\Downarrow\rangle:\ &\left\{0,0,\frac{\sqrt{6}}{5} \varepsilon,0,0,0,-1-\frac{\varepsilon}{10}-\frac{w}{2},0,0,0\right\},\label{eq:A6}
\end{align}
where we have not normalized the eigenstates and $w \equiv \sqrt{ \varepsilon^2+4 \varepsilon/5 +4}$. We have labeled these states by the diagonal matrix elements of $J_z$ and $u$ for $\varepsilon\to 0$. The label $\pm$ represents the sign of the diagonal matrix element $J_z$ in Eq.~(\ref{eq:matJz}), while the quadrupole part $\Uparrow(\Downarrow)$ in the $\Gamma_8$ states indicates the positive (negative) diagonal matrix element $u$ in Eq.~(\ref{eq:matu}).

\section{Multipole operators}\label{app:multipole}
We here list various multipole operators appearing in the Hamiltonian (\ref{eq:Hint}).
In the following, we omit the site index $i$ for simplicity. 
\begin{align}
	S_{xy}^\mu &= \frac{1}{3}\left[\frac{1}{2} \hat{\sigma}^\mu+ \frac{1}{2} \hat{\sigma}^\mu\hat{u}\right],\\
	S_{yz}^\mu &= \frac{1}{3}\left[\frac{1}{2} \hat{\sigma}^\mu+ \frac{1}{2} \hat{\sigma}^\mu\left(-\frac{1}{2}\hat{u} +\frac{\sqrt{3}}{2}\hat{v}\right)\right],\\
	S_{zx}^\mu &= \frac{1}{3}\left[\frac{1}{2} \hat{\sigma}^\mu+ \frac{1}{2} \hat{\sigma}^\mu\left(-\frac{1}{2}\hat{u} -\frac{\sqrt{3}}{2}\hat{v}\right)\right].
\end{align}
Here, $\hat{\sigma}^\mu/2$ represents the spin operator of the $t_{2g}$ orbital: $S_{xy}^\mu +S_{yz}^\mu+S_{zx}^\mu$, 
 and $\frac{1}{2}\hat{\sigma}^\mu\hat{u}$ and $\frac{1}{2}\hat{\sigma}^\mu\hat{v}$ are the spin-orbital composite ones: $2S_{xy}^\mu -S_{yz}^\mu-S_{zx}^\mu$ and  $\sqrt{3}(S_{yz}^\mu-S_{zx}^\mu)$, respectively. In terms of the irreducible representation of the O$_h$ group, the operators belonging to $\Gamma_4$  are 
 \begin{align}
 	& {\bm S}\equiv \frac{1}{2}\{\hat{\sigma}^x,\hat{\sigma}^y,\hat{\sigma}^z\},\\
 	& {\bm T}_{\alpha}\equiv \frac{1}{2}\left\{\hat{\sigma}^x\left(-\frac{1}{2}\hat{u}+\frac{\sqrt{3}}{2}\hat{v}\right),\hat{\sigma}^y\left(-\frac{1}{2}\hat{u}-\frac{\sqrt{3}}{2}\hat{v}\right),\hat{\sigma}^z\hat{u}\right\},
 \end{align}
and those belonging to $\Gamma_5$ are
 \begin{align}
 	& {\bm T}_\beta\equiv \frac{1}{2}\left\{\hat{\sigma}^x\left(-\frac{\sqrt{3}}{2}\hat{u}-\frac{1}{2}\hat{v}\right),\hat{\sigma}^y\left(\frac{\sqrt{3}}{2}\hat{u}-\frac{1}{2}\hat{v}\right),\hat{\sigma}^z\hat{v}\right\}.
 \end{align}
Using these irreps., one finds
\begin{align}
	S_{xy}^x&=\frac{1}{3}\left[S^x-\frac{1}{2}T_\alpha^x-\frac{\sqrt{3}}{2}T_\beta^x    \right],\\
	S_{xy}^y&=\frac{1}{3}\left[S^y-\frac{1}{2}T_\alpha^y+\frac{\sqrt{3}}{2}T_\beta^y    \right],\\
	S_{xy}^z&=\frac{1}{3}(S^z+T_\alpha^z),\\
	S_{yz}^x&=\frac{1}{3}(S^x+T_\alpha^x),\\
	S_{yz}^y&=\frac{1}{3}\left[S^y-\frac{1}{2}T_\alpha^y-\frac{\sqrt{3}}{2}T_\beta^y    \right]	,\\
	S_{yz}^z&=\frac{1}{3}\left[S^z-\frac{1}{2}T_\alpha^z+\frac{\sqrt{3}}{2}T_\beta^z    \right]	,\\
	S_{zx}^x&=\frac{1}{3}\left[S^x-\frac{1}{2}T_\alpha^x+\frac{\sqrt{3}}{2}T_\beta^x     \right]	,\\
	S_{zx}^y&=\frac{1}{3}(S^y+T_\alpha^y),\\
	S_{zx}^z&=\frac{1}{3}\left[S^z-\frac{1}{2}T_\alpha^z-\frac{\sqrt{3}}{2}T_\beta^z    \right]	.
\end{align}

\section{Mean-field Hamiltonian}\label{app:Hmf}
We show the detail expression of the two-site mean-field Hamiltonian used in the maintext. The total mean-field Hamiltonian is $H_{\rm int}^{\rm mf}=H_{\rm int}^{\rm A}+H_{\rm int}^{\rm B}+C$, where $H_{\rm int}^{\rm A(B)}$ is the mean-field Hamiltonian at the A(B) sublattice and $C$ is a constant depending on the order parameters. 
Denoting the expectation values of order parameters $X$ at the A(B) sublattice as $\langle X\rangle_{\rm A(B)} $, we obtain $H_{\rm int}^{\rm A}=\sum_{i\in{\rm A}}(H_{{\rm int},i}^{{\rm A},xy}+H_{{\rm int},i}^{{\rm A},yz}+H_{{\rm int},i}^{{\rm A},zx})$, where 
\begin{align}
	H_{{\rm int},i}^{{\rm A},xy}=&z_{\rm nn}\Big\{ \langle \QQ\rangle_{\rm A} \cdot \Big(\tilde{g}_{\rm iso} \mathbbm{1}+\tilde{g}_{\rm ani}  {\mathsf K}^{3} \Big) \QQ_i
	-J'\langle \bm{S}_{xy}\rangle_{\rm A} \cdot \bm{S}_{i}\nonumber\\
	&+\Big[(J+2J')
	\langle \bm{S}_{xy}\rangle_{\rm A} -J'\langle \bm{S}\rangle_{\rm A}\Big]\cdot \bm{S}_{i,xy}\Big\},\label{eq:HintAxy}\\
	H_{{\rm int},i}^{{\rm A},yz}=&z_{\rm nn}\Big\{
\langle \QQ\rangle_{\rm B} \cdot \Big(\tilde{g}_{\rm iso} \mathbbm{1}+\tilde{g}_{\rm ani}  {\mathsf K}^{1} \Big)\QQ_i
-J'\langle \bm{S}_{yz}\rangle_{\rm B} \cdot \bm{S}_{i}\nonumber\\
	&+\Big[(J+2J')
	\langle \bm{S}_{yz}\rangle_{\rm B} -J'\langle \bm{S}\rangle_{\rm B}\Big]\cdot \bm{S}_{i,yz}\Big\}, \label{eq:HintAyz}
\end{align}
where $i$ obviously belongs to the A sublattice sites and $z_{\rm nn}=4$ is the number of the nearest neighbor sites on each of the $xy, yz$, and $zx$ plane. Similarly to $H_{\rm int}^{{\rm A},yz}$, $H_{\rm int}^{{\rm A},zx}$ can be obtained by replacing $yz\to zx$ and ${\mathsf K}^1\to {\mathsf K}^2$ in Eq.~(\ref{eq:HintAyz}). The mean-field Hamiltonian at the B sublattice $H_{\rm int}^{\rm B}$ is also trivially obtained by replacing A$\leftrightarrow$B in Eqs.~(\ref{eq:HintAxy}) and (\ref{eq:HintAyz}). The constant part $C$ is necessary for calculating the free energy and this is given as
\begin{align}
	\frac{C}{z_{\rm nn}N}=& \sum_{\alpha={\rm A,B}}\Bigg[-\langle \QQ\rangle_\alpha\frac{
\tilde{g}_{\rm iso} \mathbbm{1}+\tilde{g}_{\rm ani}{\mathsf K}^{3}}{2}
\cdot \langle \QQ\rangle_\alpha \nonumber\\
	&+J'\langle \bm{S}_{xy}\rangle_\alpha \cdot \langle \bm{S}\rangle_\alpha-\frac{J+2J'}{2}
	|\langle \bm{S}_{xy}\rangle_\alpha|^2 
	\Bigg]\nonumber\\
	+&\Bigg[
	-\langle\QQ\rangle_{\rm A}\Big(
\tilde{g}_{\rm iso} \mathbbm{1}+\tilde{g}_{\rm ani}{\mathsf K}^{1}\Big)
\cdot \langle \QQ\rangle_{\rm B} \nonumber\\
&+J'\Big( \langle \bm{S}_{yz}\rangle_{\rm A} \cdot \langle \bm{S}\rangle_{\rm B}
+\langle \bm{S}\rangle_{\rm A} \cdot \langle \bm{S}_{yz}\rangle_{\rm B}\Big)
\nonumber\\
	&-(J+2J')
	\langle \bm{S}_{yz}\rangle_{\rm A}\cdot \langle \bm{S}_{yz}\rangle_{\rm B}   + (yz\to zx)	
	\Bigg].
\end{align}
Note that $N$ is the number of the unit cell (= the number of the A sublattice) and remember 
${\mathsf K}^1\to {\mathsf K}^2$ when $yz\to zx$.

\section{Matrix form of multipole operators}
The explicit 6$\times$6 matrix forms of the spin and angular momentum operators are listed below. 
First, the spin operators ${\bm S}$'s are given with $\alpha= \frac{3\sqrt{2}}{100}\varepsilon^2+O(\varepsilon^3)\simeq \sqrt{2}\delta$ in Eq.~(\ref{eq:matu}) as, 
%
%
\begin{align}
	S^x &\simeq \frac{1}{6}\small\begin{bmatrix}
		0 & 1            & 0 & \!\!\!\!-\sqrt{2}+\alpha & 0 & \!\!\!\!-\sqrt{6}+\sqrt{3}\alpha\\
		           & 0 & \sqrt{2}-\alpha & 0 & \!\!\!\!-\sqrt{6}+\sqrt{3}\alpha & 0\\	
		           &   & 0 & \!\!\!\!-2+2\sqrt{2}\alpha & 0 & \!\!\!\!\sqrt{3}-\sqrt{6}\alpha \\		
		           &   &   & 0 &\!\!\!\! -\sqrt{3}+\sqrt{6}\alpha & 0\\		
		           &             &  &  & 0 & 0\\		
		           &             &  &  &   & 0\\				
	\end{bmatrix}, \label{eq:matS1}
\end{align}
%
%
\begin{align}
	S^y &\simeq \frac{i}{6}\small\begin{bmatrix}
		0 & 1            & 0 &\!\!\!\!-\sqrt{2}+\alpha & 0 & \!\!\!\!\sqrt{6}-\sqrt{3}\alpha\\
		           & 0 & -\sqrt{2}+\alpha & 0 & \!\!\!\!-\sqrt{6}+\sqrt{3}\alpha & 0\\	
		           &   & 0 &\!\!\!\!-2+2\sqrt{2}\alpha & 0 & \!\!\!\!-\sqrt{3}+\sqrt{6}\alpha \\		
		           &   &   & 0 & \!\!\!\!-\sqrt{3}+\sqrt{6}\alpha & 0\\		
		           &             &  &  & 0 & 0\\		
		           &             &  &  &   & 0\\				
	\end{bmatrix}, \label{eq:matS2}
\end{align}
%
%
\begin{align}
	S^z &\simeq \frac{1}{6}\small\begin{bmatrix}
		1 & 0            & -2\sqrt{2}+2\alpha & 0 & 0 & 0\\
		           & -1 & 0 &\!\!\!\! -2\sqrt{2}+2\alpha  & 0 & 0\\		
		           &   & -1+\sqrt{2}\alpha & 0 & 0 & 0\\		
		           &   &  &\!\!\!\! 1-\sqrt{2}\alpha & 0 & 0\\		
		           &   &  &  &\!\!\!\! 3-3\sqrt{2}\alpha & 0\\		
		           &   &  &  &   &\!\!\!\! -3+3\sqrt{2}\alpha\\				
	\end{bmatrix}. \label{eq:matS3}
\end{align}
Here, we have not shown the left bottom part of the matrices since they are all Hermitian matrices. For the $t_{2g}$ spin operators ${\bm S}_{\rho}$'s, 
%
%
%
\begin{align}
	S_{xy}^x &\simeq \frac{1}{6}\begin{bmatrix}
		0 & 1            & 0 & -\sqrt{2}+\alpha & 0 & 0\\
		           & 0 & \sqrt{2}-\alpha & 0 & 0 & 0\\		
		           &   & 0 & -2+2\sqrt{2}\alpha & 0 & 0\\		
		           &   &   & 0 & 0 & 0\\		
		           &             &  &  & 0 & 0\\		
		           &             &  &  &   & 0\\				
	\end{bmatrix}, \label{eq:matSxy1}
\end{align}
\begin{align}
	S_{xy}^y &\simeq \frac{i}{6}\begin{bmatrix}
		0 & 1            & 0 & -\sqrt{2}+\alpha & 0 & 0\\
		           & 0 & -\sqrt{2}+\alpha & 0 & 0 & 0\\		
		           &             & 0 & -2+2\sqrt{2}\alpha & 0 & 0\\		
		          &  &  & 0 & 0 & 0\\		
		           &             &  &  & 0 & 0\\		
		           &             &  &  &  & 0\\				
	\end{bmatrix}, \label{eq:matSxy2}
\end{align}
\begin{align}
	S_{xy}^z &\simeq \frac{1}{6}\begin{bmatrix}
		-1 & 0            & -\sqrt{2}+\alpha & 0 & 0 & 0\\
		           & 1 & 0 & -\sqrt{2}+\alpha & 0 & 0\\		
		           &   & -2+2\sqrt{2}\alpha & 0 & 0 & 0\\		
		           &   &  & 2-2\sqrt{2}\alpha & 0 & 0\\		
		           &   &  &  & 0 & 0\\		
		           &   &  &  &   & 0\\				
	\end{bmatrix}, \label{eq:matSxy3}
\end{align}
%
%
%
\begin{align}
	S_{yz}^x &\simeq \frac{1}{6}\begin{bmatrix}
		0 & -1            & 0 & \frac{-1}{\sqrt{2}}+\frac{\alpha}{2} & 0 & \frac{-\sqrt{3}}{\sqrt{2}}+\frac{\sqrt{3}\alpha}{2} \\
		           & 0 & \frac{1}{\sqrt{2}}-\frac{\alpha}{2} & 0 & \frac{-\sqrt{3}}{\sqrt{2}}+\frac{\sqrt{3}\alpha}{2} & 0\\		
		           &          & 0 & \frac{1}{2}-\frac{\alpha}{\sqrt{2}} & 0 & \frac{\sqrt{3}}{2}-\frac{\sqrt{3}\alpha}{\sqrt{2}} \\		
		       &  &   & 0 & \frac{-\sqrt{3}}{2}+\frac{\sqrt{3}\alpha}{\sqrt{2}} & 0\\		
	          &          &  & & 0 & \frac{-3}{2}+\frac{3\alpha}{\sqrt{2}} \\		
	& & & &  & 0\\				
	\end{bmatrix}, \label{eq:matSyz1}
\end{align}
\begin{align}
	S_{yz}^y &\simeq \frac{i}{6}\begin{bmatrix}
		0 & 1            & 0 & \frac{1}{\sqrt{2}}-\frac{\alpha}{2} & 0 & \frac{\sqrt{3}}{\sqrt{2}}-\frac{\sqrt{3}\alpha}{2} \\
		         & 0 & \frac{1}{\sqrt{2}}-\frac{\alpha}{2} & 0 & -\frac{\sqrt{3}}{\sqrt{2}}+\frac{\sqrt{3}\alpha}{2} & 0\\		
		           &            & 0 & \frac{-1}{2}+\frac{\alpha}{\sqrt{2}} & 0 & \frac{-\sqrt{3}}{2} +\frac{\sqrt{3}\alpha}{\sqrt{2}}\\		
		        &  &  & 0 & \frac{-\sqrt{3}}{2} +\frac{\sqrt{3}\alpha}{\sqrt{2}} & 0\\		
		          &            &  &  & 0 & \frac{3}{2}-\frac{3\alpha}{\sqrt{2}}\\		
		          &           & & &  & 0\\				
	\end{bmatrix}, \label{eq:matSyz2}
\end{align}
\begin{align}
	S_{yz}^z &\simeq \frac{1}{6}\begin{bmatrix}
		1 & 0            & \frac{-1}{\sqrt{2}}+\frac{\alpha}{2} & 0 & \frac{\sqrt{3}}{\sqrt{2}}-\frac{\sqrt{3}\alpha}{2} & 0\\
	 & -1 & 0 & \frac{-1}{\sqrt{2}}+\frac{\alpha}{2} & 0 & \frac{-\sqrt{3}}{\sqrt{2}}+\frac{\sqrt{3}\alpha}{2}\\		
		  &             & \frac{1}{2}-\frac{\alpha}{\sqrt{2}} & 0 & \frac{-\sqrt{3}}{2}+\frac{\sqrt{3}\alpha}{\sqrt{2}} & 0\\		
		          & &  & \frac{-1}{2}+\frac{\alpha}{\sqrt{2}} & 0 & \frac{-\sqrt{3}}{2}+\frac{\sqrt{3}\alpha}{\sqrt{2}}\\		
		        &          & &  & \frac{3}{2}-\frac{3\alpha}{\sqrt{2}} & 0\\		
		          &  &  & &  & \frac{-3}{2}+\frac{3\alpha}{\sqrt{2}}\\				
	\end{bmatrix}, \label{eq:matSyz3}
\end{align}
%
%
\begin{align}
	S_{zx}^x &\simeq \frac{1}{6}\begin{bmatrix}
		0 & 1            & 0 & \frac{1}{\sqrt{2}}-\frac{\alpha}{2} & 0 & \frac{-\sqrt{3}}{\sqrt{2}}+\frac{\sqrt{3}\alpha}{2} \\
		           & 0 & \frac{-1}{\sqrt{2}}+\frac{\alpha}{2} & 0 & \frac{-\sqrt{3}}{\sqrt{2}}+\frac{\sqrt{3}\alpha}{2} & 0\\		
		           &          & 0 & -\frac{1}{2}+\frac{\alpha}{\sqrt{2}} & 0 & \frac{\sqrt{3}}{2}-\frac{\sqrt{3}\alpha}{\sqrt{2}} \\		
		       &  &   & 0 & \frac{-\sqrt{3}}{2}+\frac{\sqrt{3}\alpha}{\sqrt{2}} & 0\\		
	          &          &  & & 0 & \frac{3}{2}-\frac{3\alpha}{\sqrt{2}} \\		
	& & & &  & 0\\				
	\end{bmatrix}, \label{eq:matSzx1}
\end{align}
\begin{align}
	S_{zx}^y &\simeq \frac{i}{6}\begin{bmatrix}
		0 & -1            & 0 & \frac{-1}{\sqrt{2}}+\frac{\alpha}{2} & 0 & \frac{\sqrt{3}}{\sqrt{2}}-\frac{\sqrt{3}\alpha}{2} \\
		         & 0 & \frac{-1}{\sqrt{2}}+\frac{\alpha}{2} & 0 & -\frac{\sqrt{3}}{\sqrt{2}}+\frac{\sqrt{3}\alpha}{2} & 0\\		
		           &            & 0 & \frac{1}{2}-\frac{\alpha}{\sqrt{2}} & 0 & \frac{-\sqrt{3}}{2} +\frac{\sqrt{3}\alpha}{\sqrt{2}}\\		
		        &  &  & 0 & \frac{-\sqrt{3}}{2} +\frac{\sqrt{3}\alpha}{\sqrt{2}} & 0\\		
		          &            &  &  & 0 & \frac{-3}{2}+\frac{3\alpha}{\sqrt{2}}\\		
		          &           & & &  & 0\\				
	\end{bmatrix}, \label{eq:matSzx2}
\end{align}
\begin{align}
	S_{zx}^z &\simeq \frac{1}{6}\begin{bmatrix}
		1 & 0            & \frac{-1}{\sqrt{2}}+\frac{\alpha}{2} & 0 & -\frac{\sqrt{3}}{\sqrt{2}}+\frac{\sqrt{3}\alpha}{2} & 0\\
	 & -1 & 0 & \frac{-1}{\sqrt{2}}+\frac{\alpha}{2} & 0 & \frac{\sqrt{3}}{\sqrt{2}}-\frac{\sqrt{3}\alpha}{2}\\		
		  &             & \frac{1}{2}-\frac{\alpha}{\sqrt{2}} & 0 & \frac{\sqrt{3}}{2}-\frac{\sqrt{3}\alpha}{\sqrt{2}} & 0\\		
		          & &  & \frac{-1}{2}+\frac{\alpha}{\sqrt{2}} & 0 & \frac{\sqrt{3}}{2}-\frac{\sqrt{3}\alpha}{\sqrt{2}}\\		
		        &          & &  & \frac{3}{2}-\frac{3\alpha}{\sqrt{2}} & 0\\		
		          &  &  & &  & \frac{-3}{2}+\frac{3\alpha}{\sqrt{2}}\\				
	\end{bmatrix}. \label{eq:matSzx3}
\end{align}

Finally, we show the expression of the orbital angular momentum $\bm{L}$: 
\begin{align}
	L_x &\simeq \frac{1}{3}\small\begin{bmatrix}
		0 & 2 & 0 & -\frac{1}{\sqrt{2}}+\frac{\alpha_L}{2}  & 0 & -\frac{\sqrt{3}}{\sqrt{2}}+\frac{\sqrt{3}\alpha_L}{2}\\
		 & 0 &   \frac{1}{\sqrt{2}}-\frac{\alpha_L}{2}   & 0 &  -\frac{\sqrt{3}}{\sqrt{2}}+\frac{\sqrt{3}\alpha_L}{2}  & 0\\		
		 &   & 0 & 2-\eta_L  & 0 & -\sqrt{3}+\zeta_L\\		
		  &  &   & 0 & \sqrt{3}-\zeta_L  & 0\\		
		 &   &  &   & 0 & \kappa_L\\		
		  &  &   &  &   &0\\				
	\end{bmatrix}, \label{eq:matL1}
\end{align}

\begin{align}
	L_y &\simeq \frac{i}{3}\small\begin{bmatrix}
		0 & 2 & 0 & -\frac{1}{\sqrt{2}}+\frac{\alpha_L}{2}  & 0 & \frac{\sqrt{3}}{\sqrt{2}}-\frac{\sqrt{3}\alpha_L}{2}\\
		 & 0 &   -\frac{1}{\sqrt{2}}+\frac{\alpha_L}{2}   & 0 &  -\frac{\sqrt{3}}{\sqrt{2}}+\frac{\sqrt{3}\alpha_L}{2}  & 0\\		
		 &   & 0 & 2-\eta_L  & 0 & \sqrt{3}-\zeta_L\\		
		  &  &   & 0 & \sqrt{3}-\zeta_L  & 0\\		
		 &   &  &   & 0 & \kappa_L\\		
		  &  &   &  &   &0\\				
	\end{bmatrix}, \label{eq:matL2}
\end{align}
\begin{align}
	L_z &\simeq \frac{1}{3}\small\begin{bmatrix}
		2 & 0            & -\sqrt{2}+\alpha_L & 0 & 0 & 0\\
		           & -2 & 0 & -\sqrt{2}+\alpha_L & 0 & 0\\		
		           &             & 1+\beta_L & 0 & 0 & 0\\		
		           &  &  & -1-\beta_L & 0 & 0\\		
		           &             &  &  & -3-\gamma_L & 0\\		
		           &             &  &  &  & 3+\gamma_L\\				
	\end{bmatrix}, \label{eq:matL3}
\end{align}
where $\kappa_L=\frac{9}{5}\varepsilon-\frac{9}{50}\varepsilon^2+O(\varepsilon^3)$, $\eta_L=\frac{3}{5}\varepsilon+\frac{3}{50}\varepsilon^2+O(\varepsilon^3)$,
$\zeta_L=-\frac{3\sqrt{3}}{5}\varepsilon+\frac{3\sqrt{3}}{25}\varepsilon^2+O(\varepsilon^3)$, and 
 $\alpha_L= \frac{3\sqrt{2}}{5}\varepsilon-\frac{3}{50\sqrt{2}}\varepsilon^2 +O(\varepsilon^3)$, 
$\beta_L=\frac{12 }{5 }\varepsilon-\frac{3}{10}\varepsilon^2
-\frac{63}{250}\varepsilon^3+O(\varepsilon^4)$, and 
$\gamma_L=-\frac{9 }{50}\varepsilon^2
+O(\varepsilon^3)$.



\newpage

\color{black}

%

\end{document}